\documentclass[a4paper,12pt]{article}

\usepackage{graphicx}
\usepackage{amsmath}
\usepackage{amssymb}
\usepackage{hyperref}
\usepackage{tabularx}
\newcolumntype{M}{>{$}c<{$}}
\usepackage[matrix,arrow,curve]{xy}

\numberwithin{equation}{section} \numberwithin{figure}{section}
\numberwithin{table}{section}

\def\papertitlepage{\baselineskip 3.5ex\thispagestyle{empty}}
\def\Title#1{\baselineskip 1cm \vspace{1.5cm}%
  \begin{center}{\Large\bf #1}\end{center}\vspace{0.5cm}}
\def\Authors#1{\begin{center}\renewcommand{\thefootnote}{\fnsymbol{footnote}}{\it #1}\end{center}}
\def\Abstract{\vspace{1.0cm}%
  \begin{center}{\large\bf Abstract}\end{center}}

\renewenvironment{thebibliography}{\pagebreak[3]\par\vspace{0.6em}
\begin{flushleft}{\large \bf References}\end{flushleft}
\vspace{-1.0em}

\begin{enumerate}\if@twocolumn\baselineskip=0.6em\itemsep -0.2em
\else\itemsep -0.2em\fi\labelsep 0.1em}{\end{enumerate} }

\DeclareMathDelimiter{\lcolon}{\mathopen}{operators}{"3A}{largesymbols}{"3A}
\DeclareMathDelimiter{\rcolon}{\mathclose}{operators}{"3A}{largesymbols}{"3A}

\def\+{\!\!+\!\!}

\def\dynkin(#1){(#1)}

\def\bra<#1|{\langle#1|}
\def\ket|#1>{|#1\rangle}
\def\braket<#1|#2>{\langle#1|#2\rangle}
\def\llangle{\langle\!\langle}
\def\rrangle{\rangle\!\rangle}
\def\bbra<#1|{\llangle#1|}
\def\kket|#1>{|#1\rrangle}
\def\bbraket<#1|#2>{\llangle#1|#2\rrangle}

\begin{document}
{\papertitlepage \vspace*{0cm} {\hfill
\begin{minipage}{4.2cm}
IFT-P.004/2009\par\noindent May, 2009
\end{minipage}}
\Title{Cubic interaction term for Schnabl's solution using
Pad\'{e} approximants}
\Authors{{\sc E.~Aldo~Arroyo${}$\footnote{\tt
aldohep@ift.unesp.br}}
\\
${}$Instituto de F\'{i}sica Te\'{o}rica, S\~{a}o Paulo State University, \\[-2ex]
Rua Dr. Bento Teobaldo Ferraz 271, S\~{a}o Paulo, SP 01140-070,
Brasil
\\
${}$ }

} 

\vskip-\baselineskip
{\baselineskip .5cm \Abstract We evaluate the cubic interaction
term in the action of open bosonic string field theory for
Schnabl's solution written in terms of Bernoulli numbers. This
computation provides us with a new evidence for the fact that the
string field equation of motion is satisfied when it is contracted
with the solution itself.
 }
\newpage
\setcounter{footnote}{0}
\tableofcontents

\section{Introduction}

A long-standing conjecture by Sen \cite{Sen:1999mh,Sen:1999xm}
states that, at the stationary point of the tachyon potential on a
D25-brane of open bosonic string theory, the negative energy
density exactly cancels the tension of the D25-brane. The tachyon
potential in Witten's cubic open string field theory
\cite{Witten:1985cc} has been computed and numerical evidence for
Sen's conjecture was given by an approximation scheme called level
truncation
\cite{Kostelecky:1988ta,Kostelecky:1989nt,Kostelecky:1995qk,Sen:1999nx,Moeller:2000xv,Taylor:2002fy,Gaiotto:2002wy}.

The action for open bosonic string field theory is
\begin{eqnarray}
\label{accionx1} S=-\frac{1}{g^2}\Big[ \frac{1}{2} \langle
\Phi,Q_B \Phi \rangle +\frac{1}{3} \langle \Phi,\Phi*\Phi \rangle
\Big] \, ,
\end{eqnarray}
where $Q_B$ is the BRST operator of bosonic string theory,
$*$ stands for Witten's star product, and the inner product $\langle \cdot , \cdot \rangle $ is the
standard BPZ inner product. The string field $\Phi$ belongs to the
full Hilbert space of the first-quantized open string theory.

According to Sen's conjecture, the classical open string field
equation of motion
\begin{eqnarray}
\label{equamo1} Q_B \Phi + \Phi*\Phi =0 \,
\end{eqnarray}
should admit a Poincar\'{e} invariant solution $\Phi\equiv\Psi$
corresponding to the condensation of the open-string tachyon to
the vacuum with no D25-branes. This statement means that the
energy density of the true vacuum found by solving the equation of
motion should be equal to minus the tension of the D25-brane.
Since the energy density of a static configuration is minus the
action, Sen's conjecture can be summarized as follows
\begin{eqnarray}
\label{poten1} \frac{1}{g^2}\Big[\frac{1}{2} \langle \Psi,Q_B \Psi
\rangle +\frac{1}{3} \langle \Psi,\Psi*\Psi \rangle  \Big]= -
\frac{1}{2 \pi^2 g^2} \, .
\end{eqnarray}

The string field equation of motion and Sen's conjecture allow us
to fix the kinetic and cubic terms,
\begin{eqnarray}
\label{kinec1} \frac{\pi^2}{3} \langle \Psi,Q_B \Psi \rangle &=&-1 \, ,\\
\label{cubic1} \frac{\pi^2}{3} \langle \Psi,\Psi*\Psi \rangle &=&1
\, .
\end{eqnarray}

Recently, Schnabl \cite{Schnabl:2005gv} found an analytic solution
to the string field equation of motion, and it was subsequently
shown that his solution represents the nonperturbative tachyon
vacuum
\cite{Ellwood:2006ba,Ellwood:2008jh,Kawano:2008ry,Kawano:2008jv,Kiermaier:2008qu,Kishimoto:2009cz,Okawa:2006vm,Fuchs:2006hw,Takahashi:2007du}.
There are two ways of writing Schnabl's analytic solution; the
first way is in terms of Bernoulli numbers $B_n$,
\begin{eqnarray}
\label{bernu1} \Psi &=& \sum_{n,p} f_{n,p} ({\cal L}_0 + {\cal
L}_0^\dagger)^n \tilde c_p |0\rangle  + \sum_{n,p,q} f_{n,p,q}
({\cal B}_0 + {\cal B}_0^\dagger) ({\cal L}_0 + {\cal
L}_0^\dagger)^n \tilde c_p \tilde c_q  |0\rangle \, , \\
f_{n, p} &=& \frac{1-(-1)^p}{2}\frac{\pi^{-p}}{2^{n - 2 p + 1}}
\frac{1}{n!} (-1)^
      n  B_{n - p + 1} \, , \\
f_{n, p, q} &=& \frac{1-(-1)^{p + q}}{2} \frac{\pi^{-p - q}}{2^{n
- 2 (p + q) + 3}} \frac{1}{n!} (-1)^{n -q} B_{n - p - q + 2} \, ,
\end{eqnarray}
whereas the second is in terms of wedge states with ghost
insertions,
\begin{eqnarray}
\label{wedge1} \Psi &=& \lim_{N \rightarrow \infty} \Big[ \psi_N-
\sum_{n=0}^{N}
\partial_n \psi_n \Big] \; , \\
\psi_n &=& \frac{2}{\pi^2} U^\dag_{n+2}U_{n+2} \big[
(\mathcal{B}_0+\mathcal{B}^\dag_0)\tilde c(-\frac{\pi}{4}n)\tilde
c(\frac{\pi}{4}n) +\frac{\pi}{2} (\tilde c(-\frac{\pi}{4}n) +
\tilde c(\frac{\pi}{4}n)) \big] | 0\rangle \, ,
\end{eqnarray}
where $\psi_N$ with $N\rightarrow\infty$ is called the phantom
term
\cite{Okawa:2006vm,Fuchs:2006hw,Takahashi:2007du,Erler:2007xt,Aref'eva:2009ac}.

Schnabl's analytic solution was used to prove Sen's conjecture
(\ref{poten1}). Nevertheless there were subtleties involved in the
proof. For instance in a series of two subsequent papers
\cite{Okawa:2006vm,Fuchs:2006hw} it has been argued that the
validity of Schnabl's solution requires that the string field
equation of motion be satisfied when it is contracted with the
solution itself. This requirement was verified by computing the
cubic term (\ref{cubic1}) using Schnabl's solution in terms of
wedge states with ghost insertions (\ref{wedge1}). Further
numerical evidence for this result was given in
\cite{Takahashi:2007du}, where the cubic term was evaluated by
using level-truncation computations, i.e., by employing Schnabl's
solution written in the usual Virasoro basis.

In this work we use Schnabl's solution written in terms of
Bernoulli numbers (\ref{bernu1}) to provide new evidence that the
cubic term has the expected value (\ref{cubic1}) predicted from
the equation of motion and Sen's conjecture. We evaluate the cubic
term using Pad\'{e} approximants \cite{AldoArroyo:2008zm,pade}, in
analogy with the computation of the kinetic term (\ref{kinec1})
performed in \cite{Schnabl:2005gv,tedschnabl}. We confirm the
expected value of the cubic term required for the string field
equation of motion to be satisfied when contracted with the
solution itself.

This paper is organized as follows. In section 2, we evaluate the
cubic term in the action of open bosonic string field theory using
Schnabl's solution written in terms of Bernoulli numbers. Here we
use Pad\'{e} approximants to describe how to obtain the expected
value of the cubic term. A summary and further directions of
exploration are given in section 3. Some details of our
calculations such as the evaluation of correlation functions in
the ${\cal L}_0$ basis and explicit Pad\'{e} approximants
computations are given in the appendices.

\section{Evaluating the cubic term}

In this section, instead of using the representation of the
solution in terms of wedge states with ghost insertions
(\ref{wedge1}) or using the solution written in the usual Virasoro
basis \cite{Takahashi:2007du}, we evaluate the cubic term using
the solution written in terms of Bernoulli numbers (\ref{bernu1}).
The computations shown in this section are similar to those in
\cite{Schnabl:2005gv,tedschnabl}, where the kinetic term was
evaluated by using the solution written in the ${\cal L}_0$ basis,
and the expected value (\ref{kinec1}) was reproduced by means of
Pad\'{e} Approximants \cite{AldoArroyo:2008zm,pade}.

As described in \cite{Schnabl:2005gv,tedschnabl}, we start by
replacing the solution $\Psi$ with $z^{{\cal L}_0}\Psi$ in the
${\cal L}_0$ level truncation scheme, so that states in the ${\cal
L}_0$ level-expansion of the solution will acquire different
integer powers of $z$ at different levels. As we are going to see,
the parameter $z$ is needed because we need to express the cubic
term as a formal power series expansion if we want to use Pad\'{e}
approximants. After doing our calculations, we will simply set
$z=1$.

Let us start with the evaluation of the cubic term as a formal
power series expansion in $z$. Plugging the solution
(\ref{bernu1}) into the cubic term and using the correlation
functions derived in appendix A we obtain
\begin{align}
\label{expansion1} \langle \Psi ,z^{{\cal L}_0^\dag}(z^{{\cal
L}_0}\Psi)*(z^{{\cal L}_0}\Psi) \rangle & =  \frac{81 \sqrt{3}}{8
\pi^3}\frac{1}{z^3}+\Big[ -\frac{81 \sqrt{3}}{8 \pi^3} +
\frac{27}{8 \pi^2} \Big]\frac{1}{z^2} + \Big[ \frac{9 \sqrt{3}}{4
\pi^3} -\frac{3}{2 \pi^2} -\frac{\sqrt{3}}{24 \pi}
\Big]\frac{1}{z} \nonumber \\
&+ \Big[ \frac{1}{180} -\frac{13 \pi}{9720 \sqrt{3}}\Big] z +
\Big[ \frac{1}{270} -\frac{\pi}{1215 \sqrt{3}}  -
\frac{\pi^2}{21870}
\Big] z^2 \nonumber \\
&+ \Big[ \frac{5}{4536} +\frac{263 \pi}{1224720  \sqrt{3}} +
\frac{71 \pi^2}{393660}  - \frac{59 \pi^3}{8266860 \sqrt{3}}\Big]
z^3   \nonumber \\
&+ \Big[ -\frac{1}{5670} +\frac{113 \pi}{183708  \sqrt{3}} +
\frac{40 \pi^2}{137781}  - \frac{8 \pi^3}{413343 \sqrt{3}} -
\frac{5 \pi^4}{11160261}\Big]
z^4  \nonumber \\
& + \, \cdots \, .
\end{align}

At this point we remark that the most cumbersome of our
computations are the evaluation of correlation functions which
come from plugging the solution (\ref{bernu1}) into the cubic
term, the details of these computations are shown in appendix A.
Once the respective correlation functions are computed, in
principle it should be possible to write the series
(\ref{expansion1}) to any order in powers of $z$. Nevertheless,
the time it takes to do those calculations increases considerably
with every subsequent power of $z$. Given the formal power series
expansion (\ref{expansion1}), we are able to evaluate the cubic
term using Pad\'{e} approximants. We match the power series
expansion coefficients of a given rational function $P_{3+M}^N(z)$
with those of the cubic term (\ref{expansion1}). The details of
these computations can be found in appendix B.

The main result of our work is summarized in table \ref{results1}.
The first column is the definition of the cubic term in the ${\cal
L}_0$ level truncation. As we can see in the second column, the
value of the cubic term computed using Pad\'{e} approximants
converges to the expected value (\ref{cubic1}). We note that the
value of the cubic term for $n$ greater than 8 shown in the first
column has an oscillating behavior. Let us mention that a series
may diverge either by approaching infinity or by oscillating. An
example of a divergent series that diverges by going to infinity
is the series corresponding to the kinetic term
\cite{Schnabl:2005gv,tedschnabl}. It seems that in the case of the
cubic term, the divergent character of the series is due to its
oscillating behavior, which would be interesting to verify by
performing higher level computations. Since Pad\'{e} approximants
can deal numerically with divergent series
\cite{AldoArroyo:2008zm,pade} we have shown by explicit
computations that our results confirm the expected value of the
cubic term (\ref{cubic1}).

\begin{table}[ht]
\caption{The Pad\'{e} approximation for the normalized value of
the cubic term $\frac{\pi^2}{3} \langle \Psi, z^{{\cal
L}_0^\dag}(z^{{\cal L}_0}\Psi)*(z^{{\cal L}_0}\Psi) \rangle $
evaluated at $z=1$. The first column is a naive evaluation of the
cubic term given by the series (\ref{expansion1}), and the second
column is its respective $P^{n/2}_{3+n/2}$ Pad\'{e} approximation.
The label $n$ corresponds to the power of $z$ in the series
(\ref{expansion1}). At each stage of our computations we truncate
the series up to the order $z^{n-3}$.} \centering
\begin{tabular}{|c|c|c|}
\hline
  &  Naive computation  &  $P^{n/2}_{3+n/2}$ Pad\'{e} approximation     \\
    \hline
$n=0$   &  1.86073502 &  1.86073502 \\
\hline  $n=2$ & 0.96292169  & 0.91712884  \\
\hline $n=4$   & 0.97321797 & 0.97620455  \\
\hline $n=6$ & 0.98935043 & 0.97396938 \\
\hline $n=8$  & 1.00598343 & 1.00413934  \\
\hline $n=10$ & 1.00170926 & 1.00519420   \\
\hline $n=12$ & 0.99478828  & 1.00021592  \\
\hline $n=14$  & 1.00416903  & 1.00010061  \\
\hline $n=16$ & 1.00223124  & 1.00016672  \\
\hline $n=18$ & 0.99433556  & 0.99997863 \\
\hline $n=20$ & 1.00911757  & 0.99998242 \\
\hline
\end{tabular}
\label{results1}
\end{table}

\section{Summary and discussion}

We computed the cubic term in the ${\cal L}_0$ level-truncation
scheme \cite{Schnabl:2005gv,tedschnabl}, and we provided new
evidence for the fact that Schnabl's tachyon solution of open
bosonic string field theory is valid in the sense that it solves
the equation of motion when it is contracted with the solution
itself.

Up to the level that we explored with our computations, it is worth remarking
that the series that defines the cubic term
(\ref{expansion1}) seems to have an oscillating behavior. This
character of the series is in contrast with the character of the
series for the case of the kinetic term which does not begin to
diverge until higher levels, where computations reveal it starts to go to infinity
\cite{Schnabl:2005gv,tedschnabl}. In the case of the cubic term, we could perform
higher level computations to confirm the oscillating behavior of the series. We hope that the
approach used in \cite{tedschnabl} when applied to the case of the
cubic term will help to clarify this issue.

A direct application of the results shown in this paper is related
to the study of level-truncation computations in the ${\cal L}_0$
basis. In this basis, the analytic solution found by Schnabl was
originally obtained by truncating the equation of motion but not
the string field, so it would be interesting to analyze the case
when we truncate the string field instead of the equation of
motion. This analysis should serve us to address some issues, e.g., the computation of the effective tachyon potential in
Schnabl's gauge.

A second application would be the extension of our methods to the
case of the Berkovits superstring field theory
\cite{Berkovits:1995ab}. In this formalism, we already have a
solution for the tachyon condensate written in the ${\cal L}_0$
basis \cite{aldo2}. Obviously, the next step would be the evaluation
of the energy. We hope that Pad\'{e} approximants will confirm the
expected value predicted from D-brane arguments
\cite{Bagchi:2008et}.

\section*{Acknowledgements}
I would like to thank Nathan Berkovits, Ted
Erler and Martin Schnabl for useful discussions. I also wish to thank Diany Ruby,
who proofread the manuscript. This work is supported by CNPq grant 150051/2009-3.
\appendix
\setcounter{equation}{0}
\def\thesection{\Alph{section}}
\renewcommand{\theequation}{\Alph{section}.\arabic{equation}}

\section{Correlation functions and the cubic term}
All correlation functions shown in this appendix are evaluated on
the semi-infinite cylinder $C_\pi$ with circumference $\pi$. The
relation between correlation functions evaluated on the upper half
plane (UHP) and those evaluated on the semi-infinite cylinder is
given in \cite{Schnabl:2005gv}, where the conformal map $\arctan z
$ is used to map the UHP to the semi-infinite cylinder.

Employing the definition of the conformal transformation $\tilde
c(x)=\cos^2(x) c(\tan x)$ of the $c$ ghost (under the conformal
map in the paragraph above) and its anticommutator relation with
the operators $\mathcal{B}_0$ and $B_1$ \footnote{The operators
$\mathcal{B}_0$ and $B_1\equiv\mathcal{B}_{-1}$ are modes of the
$b$ ghost which are defined on the semi-infinite cylinder
coordinate as follows $\mathcal{B}_{n}=\oint \frac{dz}{2 \pi
i}(1+z^2) (\arctan z)^{n+1}b(z)$.}
\begin{align}
\{\mathcal{B}_0,\tilde c(z)\}&=z \, , \\
\{B_1,\tilde c(z)\}&=1 \, ,
\end{align}
we obtain the following basic correlation functions
\begin{align}
\label{a1}\langle \tilde{c}(x)\tilde{c}(y)\tilde{c}(z) \rangle
&=\sin(x -
y)\sin(x - z)\sin(y - z) \, , \\
\label{a2}\langle \tilde{c}(x)\mathcal{B}_0
\tilde{c}(y)\tilde{c}(z) \tilde{c}(w)\rangle &= y \langle
\tilde{c}(x)\tilde{c}(z)\tilde{c}(w)\rangle-z \langle
\tilde{c}(x)\tilde{c}(y)\tilde{c}(w)\rangle +w \langle
\tilde{c}(x)\tilde{c}(y)\tilde{c}(z) \rangle \, , \\
\label{a3}\langle \tilde{c}(x)\tilde{c}(y)\mathcal{B}_0
\tilde{c}(z) \tilde{c}(w)\rangle &= z \langle
\tilde{c}(x)\tilde{c}(y)\tilde{c}(w)\rangle -w \langle
\tilde{c}(x)\tilde{c}(y)\tilde{c}(z) \rangle \, , \\
\label{a4}\langle \tilde{c}(x)B_1 \tilde{c}(y)\tilde{c}(z)
\tilde{c}(w)\rangle &= \langle
\tilde{c}(x)\tilde{c}(z)\tilde{c}(w)\rangle-\langle
\tilde{c}(x)\tilde{c}(y)\tilde{c}(w)\rangle + \langle
\tilde{c}(x)\tilde{c}(y)\tilde{c}(z) \rangle \, , \\
\label{a5}\langle \tilde{c}(x)\tilde{c}(y)B_1 \tilde{c}(z)
\tilde{c}(w)\rangle &= \langle
\tilde{c}(x)\tilde{c}(y)\tilde{c}(w)\rangle - \langle
\tilde{c}(x)\tilde{c}(y)\tilde{c}(z) \rangle \, .
\end{align}

To compute correlation functions involved in the evaluation of the
cubic term, the following contour integrals will be very useful
\begin{align}
\label{sa}\sigma(a)&\equiv \oint \frac{dz}{2 \pi i} z^{a} \sin(2z) \nonumber \\
&= \frac{\theta(-a-2)}{\Gamma(-a)} ((-1)^{a}+1)
(-1)^{\frac{2-a}{2}} 2^{-a-2} \, , \\
\label{ca}\varsigma(a)&\equiv \oint \frac{dz}{2 \pi i} z^{a} \cos(2z) \nonumber \\
&= \frac{\theta(-a-1)}{\Gamma(-a)} ((-1)^{a}-1)
(-1)^{\frac{1-a}{2}} 2^{-a-2} \, ,
\end{align}
\begin{align}
\label{f}
\mathcal{F}(a_1,a_2,a_3,\alpha_1,\beta_1,\alpha_2,\beta_2,\alpha_3,\beta_3)\equiv
\oint \frac{dx_1 dx_2 dx_3}{(2 \pi i)^3}
x_1^{a_1}x_2^{a_2}x_3^{a_3} \langle \tilde c(\alpha_1 x_1
+\beta_1) \tilde c(\alpha_2 x_2 +\beta_2)\tilde c(\alpha_3 x_3
+\beta_3) \rangle \nonumber \;\;\;\;\;\;\;\;\;\;\\
= \frac{1}{ \alpha_1^{a_1+1}\alpha_2^{a_2+1}\alpha_3^{a_3+1}}
\Big[
\;\;\;\;\;\;\;\;\;\;\;\;\;\;\;\;\;\;\;\;\;\;\;\;\;\;\;\;\;\;\;\;\;\;\;\;\;\;\;\;\;\;\;\;\;\;\;\;\;\;\;\;\;\;
 \;\;\;\;\;\;\;\;\;\;\;\;\;\;\;\;\;\;\;\;\;\;\;\;   \nonumber\\
 \delta_{a_3,-1}\frac{ \big(\sigma(a_1)\sigma(a_2)+\varsigma(a_1)\varsigma(a_2)\big) \sin(2(\beta_1-\beta_2))
 +
 \big(\sigma(a_1)\varsigma(a_2)-\varsigma(a_1)\sigma(a_2)\big) \cos(2(\beta_1-\beta_2))}{4} \nonumber \;\;\;\;\;\;\;\;\;\;\;\;\;\;\;\;\;\;\;\;\\
 + \delta_{a_2,-1}\frac{ \big(\varsigma(a_1)\sigma(a_3)-\sigma(a_1)\varsigma(a_3)\big) \cos(2(\beta_1-\beta_3))
 -
 \big(\varsigma(a_1)\varsigma(a_3)+\sigma(a_1)\sigma(a_3)\big) \sin(2(\beta_1-\beta_3))}{4} \nonumber \;\;\;\;\;\;\;\;\;\;\;\;\;\;\;\;\;\;\\
 + \delta_{a_1,-1}\frac{ \big(\varsigma(a_2)\mathcal{C}(a_3)+\sigma(a_2)\sigma(a_3)\big) \sin(2(\beta_2-\beta_3))
 +
 \big(\sigma(a_2)\varsigma(a_3)-\varsigma(a_2)\sigma(a_3)\big)
 \cos(2(\beta_2-\beta_3))}{4}
 \Big], \;\;\;\;\;\;\;\;\;\;\;\;\;\;\; \nonumber \\
\end{align}
where $\theta(n)$ is the unit step (Heaviside) function which is
defined as follows
\begin{align}
\theta(n) =
\begin{cases}
  0,  & \mbox{if }n < 0 \\
  1, & \mbox{if }n \geq 0 \, .
\end{cases}
\end{align}

Let us list a few non-trivial correlation functions which involve
operators frequently used in the $\mathcal{L}_0$ basis, namely
$\hat{\mathcal{L}}^{n}$ ($\hat{\mathcal{L}}\equiv {\cal L}_0+{\cal
L}_0^\dag$), $\hat{\mathcal{B}}$ ($\hat{\mathcal{B}}\equiv {\cal
B}_0+{\cal B}_0^\dag$), $U_r=\big(\frac{2}{r}\big)^{{\cal L}_0}$
and the  $\tilde c(z)$ ghost
\begin{align}
\label{correla1}&\langle \text{bpz}(\tilde{c}_{p_1})
\hat{\mathcal{L}}^{n_1} U^\dag_{r} U_{r}
\tilde{c}(x)\tilde{c}(y) \rangle = \nonumber \\
&= \oint \frac{dz_1 dx_1}{(2 \pi i)^{2}}\frac{(-2)^{n_1} n_1! \,
x_1^{p_1-2}}{(z_1-2)^{n_1+1}} \big(\frac{2}{r}\big)^{-p_1+n_1-2}
\big(\frac{2}{z_1}\big)^{-p_1-2} \langle
\tilde{c}(x_1+\frac{\pi}{2}) \tilde{c}(\frac{4}{z_1
r}x)\tilde{c}(\frac{4}{z_1 r}y) \rangle \, ,
\end{align}
\begin{align}
\label{correla2}&\langle \text{bpz}(\tilde{c}_{p_1})
\hat{\mathcal{L}}^{n_1} \hat{\mathcal{B}} U^\dag_{r} U_{r}
\tilde{c}(x)\tilde{c}(y)\tilde{c}(z) \rangle = \nonumber \\
&= - \delta_{p_1,0} \oint \frac{dz_1}{ 2 \pi i} \frac{(-2)^{n_1}
n_1!}{(z_1-2)^{n_1+1}} \big(\frac{2}{r}\big)^{-p_1+n_1-2}
\big(\frac{2}{z_1}\big)^{-p_1-2} \langle \tilde{c}(\frac{4}{z_1
r}x)\tilde{c}(\frac{4}{z_1 r}y)\tilde{c}(\frac{4}{z_1 r}z) \rangle
\nonumber \\
&+  \oint \frac{dz_1 dx_1}{(2 \pi i)^{2}}\frac{(-2)^{n_1} n_1! \,
x_1^{p_1-2}}{(z_1-2)^{n_1+1}} \big(\frac{2}{r}\big)^{-p_1+n_1-2}
\big(\frac{2}{z_1}\big)^{-p_1-2} \langle
\tilde{c}(x_1+\frac{\pi}{2}) \mathcal{B}_0 \tilde{c}(\frac{4}{z_1
r}x)\tilde{c}(\frac{4}{z_1 r}y)\tilde{c}(\frac{4}{z_1 r}z) \rangle
\, ,
\end{align}
\begin{align}
\label{correla3}&\langle \text{bpz}(\tilde{c}_{p_1})
\text{bpz}(\tilde{c}_{p_2}) \hat{\mathcal{L}}^{n_1}
\hat{\mathcal{B}} U^\dag_{r} U_{r}
\tilde{c}(x)\tilde{c}(y) \rangle = \nonumber \\
&= - \delta_{p_2,0}\oint \frac{dz_1 dx_1}{(2 \pi
i)^{2}}\frac{(-2)^{n_1} n_1! \, x_1^{p_1-2}}{(z_1-2)^{n_1+1}}
\big(\frac{2}{r}\big)^{-p_1-p_2+n_1-1}
\big(\frac{2}{z_1}\big)^{-p_1-p_2-1} \langle
\tilde{c}(x_1+\frac{\pi}{2}) \tilde{c}(\frac{4}{z_1
r}x)\tilde{c}(\frac{4}{z_1 r}y) \rangle \nonumber \\
& +\delta_{p_1,0}\oint \frac{dz_1 dx_2}{(2 \pi
i)^{2}}\frac{(-2)^{n_1} n_1! \, x_2^{p_2-2}}{(z_1-2)^{n_1+1}}
\big(\frac{2}{r}\big)^{-p_1-p_2+n_1-1}
\big(\frac{2}{z_1}\big)^{-p_1-p_2-1} \langle
\tilde{c}(x_2+\frac{\pi}{2}) \tilde{c}(\frac{4}{z_1
r}x)\tilde{c}(\frac{4}{z_1 r}y) \rangle \nonumber \\
& +\oint \frac{dz_1 dx_1 dx_2}{(2 \pi i)^{3}}\frac{(-2)^{n_1} n_1!
\, x_1^{p_1-2} x_2^{p_2-2}}{(z_1-2)^{n_1+1}}
\big(\frac{2}{r}\big)^{-p_1-p_2+n_1-1}
\big(\frac{2}{z_1}\big)^{-p_1-p_2-1} \times \nonumber \\
 &\;\;\;\;\;\;\;\;\;\;\;\;\;\;\;\;\;\;\;\;\;\;\;\;\;\;\;\;
 \;\;\;\;\;\;\;\;\;\;\;\;\;\;\;\;\;\;\;\;\;\;\;
 \;\;\;\;\;\;\;\;\;\;\;\;\;\;\;\;\;\; \times \langle \tilde{c}(x_1+\frac{\pi}{2})
\tilde{c}(x_2+\frac{\pi}{2}) \mathcal{B}_0 \tilde{c}(\frac{4}{z_1
r}x)\tilde{c}(\frac{4}{z_1 r}y) \rangle \, ,
\end{align}
where the ``\emph{bpz}'' acting on the modes of the $\tilde c(z)$ ghost
stands for the usual BPZ conjugation which in the ${\cal L}_0$
basis is defined as follows
\begin{align}
\text{bpz}(\tilde \phi_n) &= \oint \frac{d  z}{2 \pi i} z^{n+h-1}
\tilde \phi ( z + \frac{\pi}{2}) \, ,
\end{align}
for any primary field $\tilde \phi(z)$ with weight $h$. The action
of the BPZ conjugation on the modes of $\tilde \phi(z)$ satisfies
the following useful property
\begin{align}
U_r^{\dag -1} \text{bpz}(\tilde \phi_n)
U_r^{\dag}=\big(\frac{2}{r}\big)^{-n} \text{bpz}(\tilde \phi_n) \,
.
\end{align}

Here we provide a few intermediate steps regarding to the
evaluation of the cubic term of the open string field action for
Schnabl's solution $\Psi$ expressed in terms of Bernoulli numbers
(\ref{bernu1}). As discussed in Section 2, in order to apply
Pad\'{e} approximants we must start by replacing the solution
$\Psi$ with $z^{{\cal L}_0}\Psi$. In the ${\cal L}_0$ level
expansion different levels will acquire different integer powers
of $z$. Plugging this redefinition of the solution into the cubic
term of the action we obtain
\begin{align}
\label{cubic22} &\langle \Psi, z^{{\cal L}_0^\dag}(z^{{\cal
L}_0}\Psi)*(z^{{\cal L}_0}\Psi) \rangle  = \nonumber \\
&\sum_{n_1,n_2,n_3,p_1,p_2,p_3} f_{n_1,p_1}f_{n_2,p_2}f_{n_3,p_3}
\Delta^{(1)}_{n_1,n_2,n_3,p_1,p_2,p_3}
z^{n_1+n_2+n_3-p_1-p_2-p_3} + \nonumber \\
& \sum_{n_1,n_2,n_3,p_1,p_2,p_3,p_4}
f_{n_1,p_1}f_{n_2,p_2}f_{n_3,p_3,p_4} \Delta^{(2)}_{n_1, n_2, n_3,
p_1,p_2, p_3, p_4}z^{n_1 + n_2 + n_3 + 1 - p_1 - p_2 - p_3 - p_4} + \nonumber \\
& \sum_{n_1,n_2,n_3,p_1,p_2,p_3,p_4} f_{n_1, p_1}f_{n_2, p_2,
p_3}f_{n_3, p_4}\Delta^{(3)}_{n_1, n_2, n_3, p_1, p_2, p_3,p_4}z^{n_1 + n_2 + n_3 + 1 - p_1 - p_2 - p_3 - p_4} + \nonumber \\
& \sum_{n_1,n_2,n_3,p_1,p_2,p_3,p_4} f_{n_1, p_1, p_2}f_{n_2,
p_3}f_{n_3, p_4}
\Delta^{(4)}_{n_1, n_2, n_3, p_1, p_2, p_3,p_4}z^{n_1 + n_2 + n_3 + 1 - p_1 - p_2 - p_3 - p_4} + \nonumber \\
& \sum_{n_1,n_2,n_3,p_1,p_2,p_3,p_4,p_5} f_{n_1, p_1}f_{n_2, p_2,
p_3}f_{n_3, p_4, p_5}\Delta^{(5)}_{n_1, n_2, n_3, p_1, p_2, p_3,
p_4,p_5}z^{n_1 + n_2 + n_3 + 2 - p_1 - p_2 - p_3 - p_4 - p_5} + \nonumber \\
& \sum_{n_1,n_2,n_3,p_1,p_2,p_3,p_4,p_5} f_{n_1, p_1, p_2}f_{n_2,
p_3}f_{n_3, p_4, p_5}\Delta^{(6)}_{n_1, n_2, n_3, p_1, p_2, p_3,
p_4,p_5}z^{n_1 + n_2 + n_3 + 2 - p_1 - p_2 - p_3 - p_4 - p_5} + \nonumber \\
& \sum_{n_1,n_2,n_3,p_1,p_2,p_3,p_4,p_5} f_{n_1, p_1, p_2}f_{n_2,
p_3, p_4}f_{n_3, p_5}\Delta^{(7)}_{n_1, n_2, n_3, p_1, p_2, p_3,
p_4,p_5}z^{n_1 + n_2 + n_3 + 2 - p_1 - p_2 - p_3 - p_4 - p_5} + \nonumber \\
& \sum_{n_1,n_2,n_3,p_1,p_2,p_3,p_4,p_5,p_6} f_{n_1,
p_1,p_2}f_{n_2, p_3, p_4}f_{n_3, p_5, p_6}\Delta^{(8)}_{n_1, n_2,
n_3, p_1, p_2, p_3,
p_4,p_5,p_6}z^{n_1 + n_2 + n_3 + 3 - p_1 - p_2 - p_3 - p_4 - p_5-p_6}, \nonumber \\
\end{align}
where to simplify notation we have used the following
definitions
\begin{align}
\label{co01} \Delta^{(1)}_{n_1,n_2,n_3,p_1,p_2,p_3} &
\equiv\langle 0|\text{bpz}(\tilde c_{p_1}) \hat{\mathcal{L}}^{n_1}
,\hat{\mathcal{L}}^{n_2}  \tilde c_{p_2} |0 \rangle *
\hat{\mathcal{L}}^{n_3}  \tilde c_{p_3}|0 \rangle, \\
\label{co02} \Delta^{(2)}_{n_1,n_2,n_3,p_1,p_2,p_3,p_4} &
\equiv\langle 0|\text{bpz}(\tilde c_{p_1}) \hat{\mathcal{L}}^{n_1}
,\hat{\mathcal{L}}^{n_2}  \tilde c_{p_2} |0 \rangle *
\hat{\mathcal{L}}^{n_3} \hat{\mathcal{B}}   \tilde c_{p_3} \tilde
c_{p_4}|0 \rangle, \\
\Delta^{(3)}_{n_1,n_2,n_3,p_1,p_2,p_3,p_4} & \equiv\langle
0|\text{bpz}(\tilde c_{p_1}) \hat{\mathcal{L}}^{n_1}
,\hat{\mathcal{L}}^{n_2}\hat{\mathcal{B}}   \tilde c_{p_2} \tilde
c_{p_3}|0 \rangle * \hat{\mathcal{L}}^{n_3}  \tilde c_{p_4} |0 \rangle,\\
\Delta^{(4)}_{n_1,n_2,n_3,p_1,p_2,p_3,p_4} & \equiv\langle
0|\text{bpz}(\tilde c_{p_1})\text{bpz}(\tilde c_{p_2})
\hat{\mathcal{L}}^{n_1}\hat{\mathcal{B}} ,\hat{\mathcal{L}}^{n_2}
\tilde c_{p_3} |0 \rangle * \hat{\mathcal{L}}^{n_3} \tilde
c_{p_4}|0 \rangle,\\
\Delta^{(5)}_{n_1,n_2,n_3,p_1,p_2,p_3,p_4,p_5} & \equiv\langle
0|\text{bpz}(\tilde c_{p_1}) \hat{\mathcal{L}}^{n_1}
,\hat{\mathcal{L}}^{n_2} \hat{\mathcal{B}}\tilde c_{p_2} \tilde
c_{p_3} |0 \rangle * \hat{\mathcal{L}}^{n_3} \hat{\mathcal{B}}
\tilde c_{p_4} \tilde c_{p_5}|0 \rangle,\\
\Delta^{(6)}_{n_1,n_2,n_3,p_1,p_2,p_3,p_4,p_5} & \equiv\langle
0|\text{bpz}(\tilde c_{p_1})\text{bpz}(\tilde c_{p_2})
\hat{\mathcal{L}}^{n_1}\hat{\mathcal{B}} ,\hat{\mathcal{L}}^{n_2}
\tilde c_{p_3} |0 \rangle * \hat{\mathcal{L}}^{n_3}
\hat{\mathcal{B}} \tilde c_{p_4} \tilde c_{p_5}|0 \rangle,\\
\Delta^{(7)}_{n_1,n_2,n_3,p_1,p_2,p_3,p_4,p_5} & \equiv\langle
0|\text{bpz}(\tilde c_{p_1})\text{bpz}(\tilde c_{p_2})
\hat{\mathcal{L}}^{n_1}\hat{\mathcal{B}}
,\hat{\mathcal{L}}^{n_2}\hat{\mathcal{B}} \tilde c_{p_3} \tilde
c_{p_4}|0 \rangle * \hat{\mathcal{L}}^{n_3} \tilde c_{p_5}|0
\rangle,\\
\Delta^{(8)}_{n_1,n_2,n_3,p_1,p_2,p_3,p_4,p_5,p_6} & \equiv\langle
0|\text{bpz}(\tilde c_{p_1})\text{bpz}(\tilde c_{p_2})
\hat{\mathcal{L}}^{n_1}\hat{\mathcal{B}}
,\hat{\mathcal{L}}^{n_2}\hat{\mathcal{B}} \tilde c_{p_3} \tilde
c_{p_4}|0 \rangle * \hat{\mathcal{L}}^{n_3} \hat{\mathcal{B}}
\tilde c_{p_5}\tilde c_{p_6}|0 \rangle \,
\end{align}
for all the correlation functions appearing in the evaluation of
the cubic term.

All these correlation functions can be readily computed using the
results of this Appendix. For instance, let us compute the
correlator $\langle 0|\text{bpz}(\tilde c_{p_1})
\hat{\mathcal{L}}^{n_1} ,\hat{\mathcal{L}}^{n_2}  \tilde c_{p_2}
|0 \rangle * \hat{\mathcal{L}}^{n_3}  \tilde c_{p_3}|0 \rangle$
which involves states of the form $\hat{\mathcal{L}}^{n} \tilde
c_p |0 \rangle$
\begin{align}
&\label{corre7} \langle 0|\text{bpz}(\tilde c_{p_1})
\hat{\mathcal{L}}^{n_1} ,\hat{\mathcal{L}}^{n_2}  \tilde c_{p_2}
|0 \rangle * \hat{\mathcal{L}}^{n_3}  \tilde c_{p_3}|0 \rangle =
\nonumber \\
&= \frac{(-2)^{n_2+n_3} n_2! n_3! }{(2 \pi i)^{4}} \oint
\frac{dz_2 dz_3 dx_2 dx_3 \,
x_2^{p_2-2}x_3^{p_3-2}}{(z_2-2)^{n_2+1}(z_3-2)^{n_3+1}} \langle
0|\text{bpz}(\tilde c_{p_1}) \hat{\mathcal{L}}^{n_1}, U^\dag_{z_2}
U_{z_2} \tilde c(x_2)|0 \rangle * U^\dag_{z_3} U_{z_3} \tilde
c(x_3) |0 \rangle
\nonumber \\
&= \frac{(-2)^{n_2+n_3} n_2! n_3! }{(2 \pi i)^{4}} \oint
\frac{dz_2 dz_3 dx_2 dx_3 \,
x_2^{p_2-2}x_3^{p_3-2}}{(z_2-2)^{n_2+1}(z_3-2)^{n_3+1}} \times
\nonumber
\\ &\times \langle \text{bpz}(\tilde c_{p_1}) \hat{\mathcal{L}}^{n_1}
U^\dag_{r} U_{r} \tilde c(x_2+\frac{\pi}{4}(z_3-1))
\tilde c(x_3-\frac{\pi}{4}(z_2-1)) \rangle \nonumber \\
&=\frac{(-1)^{n_1+n_2+n_3} 2^{2 n_1+n_2+n_3-2p_1-4} n_1!n_2! n_3!
}{(2 \pi i)^{3}} \oint \frac{dz_1dz_2 dz_3 \,
z_1^{p_1+2}r^{p_1+2-n_1}}{(z_1-2)^{n_1+1}(z_2-2)^{n_2+1}(z_3-2)^{n_3+1}}
\times \nonumber
\\ &\times \mathcal{F}(p_1-2,p_2-2,p_3-2,1,\frac{\pi}{2},\frac{4}{z_1r},\frac{\pi(z_3-1)}{z_1r},
\frac{4}{z_1r},\frac{\pi(1-z_2)}{z_1r}) \, ,
\end{align}
where we have defined $r\equiv z_2+z_3-1$ and used the definition
of $\mathcal{F}$ (\ref{f}).

Although the expression for the cubic term (\ref{cubic22}) looks
complicated, it is actually quite easy to simplify the expression.
Using the cyclicity symmetry of the three vertex
\begin{eqnarray}
\langle A,B*C \rangle =
(-1)^{\text{gh}(A)(\text{gh}(B)+\text{gh}(C))}\langle B,C*A
\rangle = (-1)^{\text{gh}(C)(\text{gh}(A)+\text{gh}(B))}\langle
C,A*B \rangle \, , \nonumber
\end{eqnarray}
and the following star product identities involving the
$\mathcal{B}_0$ operator
\begin{eqnarray}
((\mathcal{B}_0+\mathcal{B}^\dag_0)\phi_1)*\phi_2 &=&
(\mathcal{B}_0+\mathcal{B}^\dag_0)(\phi_1*\phi_2) +
(-1)^{\text{gh}(\phi_1)} \frac{\pi}{2} \phi_1 * B_1 \phi_2 \; , \nonumber \\
\phi_1 * ((\mathcal{B}_0+\mathcal{B}^\dag_0)\phi_2)&=&
(-1)^{\text{gh}(\phi_1)}(\mathcal{B}_0+\mathcal{B}^\dag_0)(\phi_1*\phi_2)-(-1)^{\text{gh}(\phi_1)}\frac{\pi}{2}(B_1\phi_1)*\phi_2
\; , \nonumber \\
((\mathcal{B}_0+\mathcal{B}^\dag_0)\phi_1)*((\mathcal{B}_0+\mathcal{B}^\dag_0)\phi_2)&=&-(-1)^{\text{gh}(\phi_1)}
\frac{\pi}{2}(\mathcal{B}_0+\mathcal{B}^\dag_0)B_1(\phi_1*\phi_2)+(\frac{\pi}{2})^2
(B_1 \phi_1)*(B_1 \phi_2)\, , \nonumber
\end{eqnarray}
we obtain the following simplified expression for the cubic term
\begin{eqnarray}
\langle \Psi, z^{{\cal L}_0^\dag}(z^{{\cal L}_0}\Psi)*(z^{{\cal
L}_0}\Psi) \rangle  = \;\;\;\;\;\; \;\;\;\;\;\; \;\;\;\;\;\;
\;\;\;\;\;\;
 \;\;\;\;\;\; \;\;\;\;\;\; \;\;\;\;\;\; \;\;\;\;\;\;  \;\;\;\;\;\; \;\;\;\;\;\; \;\;\;\;\;\;
  \;\;\;\;\;\; \;\;\;\;\;\; \;\;\;\;\;\; \;\;\;\;\;\; \;\;\;\;\;\;\nonumber \\
\sum_{n_1,n_2,n_3,p_1,p_2,p_3} \big[f_{n_1, p_1}f_{n_2,
p_2}f_{n_3, p_3} + 3\pi^2 f_{n_1, p_1} f_{n_2, 1, p_2} f_{n_3, 1,
p_3} \big] \Delta^{(1)}_{n_1,n_2,n_3,p_1,p_2,p_3}
z^{n_1+n_2+n_3-p_1-p_2-p_3} + \nonumber \\
 \sum_{n_1,n_2,n_3,p_1,p_2,p_3,p_4} \big[3f_{n_1, p_1}f_{n_2,
p_2}f_{n_3, p_3, p_4} + \pi^2 f_{n_1, 1, p_1}f_{n_2, 1,
p_2}f_{n_3, p_3, p_4} + 3\pi f_{n_1, p_1}f_{n_2, 1, p_2}f_{n_3,
p_3, p_4} \nonumber  \;\;\;\;\;\; \;\;\;\;\;\\  -
    3\pi f_{n_2, p_2}f_{n_1, 1, p_1}f_{n_3, p_3, p_4}\big] \Delta^{(2)}_{n_1, n_2, n_3,
p_1,p_2, p_3, p_4}z^{n_1 + n_2 + n_3 + 1 - p_1 - p_2 - p_3 - p_4}
\, . \;\;\;\;\;\; \;\;\;\;\;\; \;\;\;\;
 \nonumber
\end{eqnarray}

Therefore we only need to compute the correlation functions
(\ref{co01}) and (\ref{co02}). As it was already mentioned these
correlation functions can be readily computed by using the
correlators (\ref{a1}), (\ref{a2}), (\ref{a3}), (\ref{a4}),
(\ref{a5}), (\ref{correla1}), (\ref{correla2}) and
(\ref{correla3}). We were aided by a computer to perform these calculations.

\section{Pad\'{e} approximant computations}
Here we shall explain the method to calculate the
cubic term based on Pad\'{e} approximants by computing in detail
the normalized value of the cubic term at order $n=4$, shown in
table \ref{results1}. At this order, we need to consider terms in
the series (\ref{expansion1}) up to linear order in $z$, namely
\begin{eqnarray}
\label{ss1} \frac{81 \sqrt{3}}{8 \pi^3}\frac{1}{z^3}+\Big[
-\frac{81 \sqrt{3}}{8 \pi^3} + \frac{27}{8 \pi^2}
\Big]\frac{1}{z^2} + \Big[ \frac{9 \sqrt{3}}{4 \pi^3} -\frac{3}{2
\pi^2} -\frac{\sqrt{3}}{24 \pi} \Big]\frac{1}{z} + \Big[
\frac{1}{180} -\frac{13 \pi}{9720 \sqrt{3}}\Big] z.  \;\;
\end{eqnarray}

Using Pad\'{e} approximants, we express (\ref{ss1}) as the following rational function
\begin{eqnarray}
\label{ss2} P^2_{3+2}(z)=\frac{1}{z^3} \Big[\frac{a_0+a_1z+a_2z^2
}{1+b_1z+b_2z^2} \Big]\, .
\end{eqnarray}
Expanding the right hand side of (\ref{ss2}) around $z=0$, we get
up to linear order in $z$
\begin{eqnarray}
\label{ss3} P^2_{3+2}(z)&=&\frac{a_0}{z^3} +
\frac{a_1-a_0b_1}{z^2}+\frac{a_2 - a_1b_1 + a_0 b_1^2 - a_0
b_2}{z} \nonumber \\
&+&( a_1b_1^2 -a_2b_1 - a_0b_1^3 - a_1b_2 +
      2a_0b_1b_2) \nonumber \\
&+&(a_2b_1^2 - a_1 b_1^3 + a_0b_1^4 - a_2 b_2 +
          2 a_1b_1 b_2 - 3 a_0 b_1^2 b_2 +
          a_0b_2^2) z \, .
\end{eqnarray}

Equating the coefficients of $z^{-3}$, $z^{-2}$, $z^{-1}$,
$z^{0}$, $z^{1}$ in equations (\ref{ss1}) and (\ref{ss3}), we get
a system of five algebraic equations for the unknown coefficients
$a_0$, $a_1$, $a_2$, $b_1$ and $b_2$. Solving these equations we
get
\begin{eqnarray}
\label{aaa1}a_0&=&0.565595624636 \, , \\
\label{aaa2}a_1&=&-0.38673808434 \, , \\
\label{aaa3}a_2&=& 0.051154789816 \, , \\
\label{aaa4}b_1&=& -0.28837113902 \, , \\
\label{aaa5}b_2&=& 0.063526545755 \, .
\end{eqnarray}

Replacing the value of the coefficients (\ref{aaa1}),
(\ref{aaa2}), (\ref{aaa3}), (\ref{aaa4}) and (\ref{aaa5}) into the
definition of $P^2_{3+2}(z)$ (\ref{ss2}), and evaluating this at
$z=1$, we get the following normalized value for the cubic term,
\begin{eqnarray}
\frac{\pi^2}{3} P^2_{3+2}(z=1) = 0.976204550211 \, .
\end{eqnarray}

\newpage

\end{document}